\documentclass[12pt]{iopart}
\usepackage{graphicx}
\input amssym.def \input amssym

\begin{document}

\title[]{Clifford quantum computer and the Mathieu groups}

\author{Michel Planat }

\address{Institut FEMTO-ST, CNRS, D\' epartement LPMO, 32 Avenue de
l'Observatoire\\ F-25044 Besan\c con, France }
\ead{michel.planat@femto-st.fr}
\begin{abstract}
One learned from Gottesman-Knill theorem that the Clifford model of quantum computing \cite{Clark07} may be generated from a few quantum gates, the Hadamard, $\pi/4$-Phase and Controlled-Z gates, and efficiently simulated on a classical computer. We employ the group theoretical package GAP\cite{GAP} for simulating the two qubit Clifford group $\mathcal{C}_2$. We already found that the symmetric group $S(6)$, aka the automorphism group of the generalized quadrangle $W(2)$, controls the geometry of the two-qubit Pauli graph \cite{Pauligraphs}. Sixfold symmetry is also revealed in the {\it inner} Clifford group  $\mbox{Inn}(\mathcal{C}_2)=\mathcal{C}_2/\mbox{Center}(\mathcal{C}_2)$. It contains two normal subgroups, one isomorphic to $\mathcal{Z}_2^{\times 4}$, and the second isomorphic to the semi-direct product $U_6=\mathcal{Z}_2^{\times 4} \rtimes A(6)$ (of order $5760$). The group $U(6)$ stabilizes an {\it hexad} in the Steiner system $S(3,6,22)$ attached to the Mathieu group $M(22)$. Both groups $A(6)$ and $U_6$ have an {\it outer} automorphism group $\mathcal{Z}_2\times \mathcal{Z}_2$, a feature we associate to two-qubit quantum entanglement.
 
\end{abstract}

\maketitle

\noindent
\hrulefill
\section{Introduction}

The Clifford model of quantum computation deals with the unitary time evolution of Pauli operators going through a selected set of generating quantum gates. 
One first notices that the traditional picture of unitary evolution of quantum states under an Hamiltonian has to be replaced by an evolution of 
operators under a properly chosen group action. This is analogous to the Heisenberg representation of quantum mechanics, where the operators evolve, as opposed to the Schr\"{o}dinger picture, where the states evolve \cite{Go98}. Pauli operators are defined as $n$-fold tensor products of the identity and the ordinary Pauli matrices. Various quantum error correcting codes have been developed for the states arising from the Pauli operators \cite{Ca97}-\cite{Kla2}. One of the challenges has been to identify the types of quantum computations that can be classically simulated. One earliest result in this context is Gottesman-Knill theorem \cite{Nielsen} which essentially relies on a stabilizer formalism and describes the unitary dynamics and the measurements. The generating quantum gates of the stabilizer formalism are the Hadamard operation $H=1/\sqrt{2}[(1,1),(1,-1)]$ and the $\pi/4$ phase gate $P=\mbox{diag}(1,i)$ acting on a single qubit, and the controlled-$Z$ gate $CZ=\mbox{diag}(1,1,1,-1)$ acting on two qubits. Using these building gates, tensor products of them and measurements of the Pauli operators, any quantum computation is, in principle, efficiently simulated on a classical computer. Another type of quantum computation, using measurements as computational steps, is being developed \cite{Raussen01}-\cite{Danos}, but it will not be the topic of the present paper.

Instead we will propose a detailed group theoretical analysis of the Clifford group acting on one or two qubits, using the group theoretical package GAP4 \cite{GAP} as our classical simulator. A number of recent papers deal with the Clifford group analysis in relation to quantum computing \cite{Clark07}, quantum measurements \cite{Appleby} or local encoding of classical information \cite{Markham}. So far the connection to the geometry of Mathieu groups seems to have remained unoticed.

Let us start with the standard single qubit $\sigma_x$ (shift), $\sigma_z$ (clock) and $\sigma_y=i \sigma_x \sigma_z$ Pauli matrices. The corresponding Pauli group $\mathcal{P}_1$ may be defined from three generators, i.e. $\mathcal{P}_1=\left\langle \sigma_x, \sigma_y, \sigma_z \right\rangle$. The Pauli group on $n$ qubits is the $n$-fold tensor product of $\mathcal{P}_1$ and has order $|\mathcal{P}_n|=2^{2n+2}$. All elements of $\mathcal{P}_n$ commuting with any other operator are in the center $Z(\mathcal{P}_n)=\{\pm 1, \pm iI\}$. 

In our previous papers we extensively studied the commutation relations within the two-qubit Pauli group \cite{Pauligraphs} and found the finite projective geometry underlying them. Skipping the elements in the center, the fifteen tensor products of Pauli matrices $\sigma_i= (I_2,\sigma_x, \sigma_y,\sigma_z)$, $\sigma_i \otimes \sigma_j$, $i,j \in \{1,2,3,4\}$ and $(i,j)\neq (1,1)$, can be labeled as $1=I_2 \otimes \sigma_x$, $2=I_2 \otimes \sigma_y$, $3=I_2 \otimes \sigma_z$, $a=\sigma_x \otimes I_2$, $4=\sigma_x \otimes \sigma_x$\ldots, $b=\sigma_y \otimes I_2$,\ldots , $c=\sigma_z \otimes I_2$,\ldots. The commutation rules between them are displayed in Fig. 1. Each {\it line} of mutually commuting operators contains exactly three {\it points} and each {\it point} is on three {\it lines}. Any point not on a line (an antiflag) is on a unique line intersecting it. This type of geometry is known as a generalized quadrangle and the two-qubit system provides the smallest example, having fifteen points and, by duality, fifteen lines. The three operators on a line share (stabilize) a common set of four quantum states known as a base. One clearly observes in Fig. 1 that a maximum of five non-intersecting lines can be obtained, which corresponds to a maximal set of five mutually unbiased bases in this dimension (see \cite{Pauligraphs},\cite{PlanatMUBs} for more about mutually unbiased bases). Six of the (boldfaced) lines form a grid of entangled bases known as a Mermin square in reference to their ability to prove the Kochen-Specker theorem in dimension four \cite{Pauligraphs,Mermin}. 

\begin{figure}[ht]
\centerline{\includegraphics[width=7.0truecm,clip=]{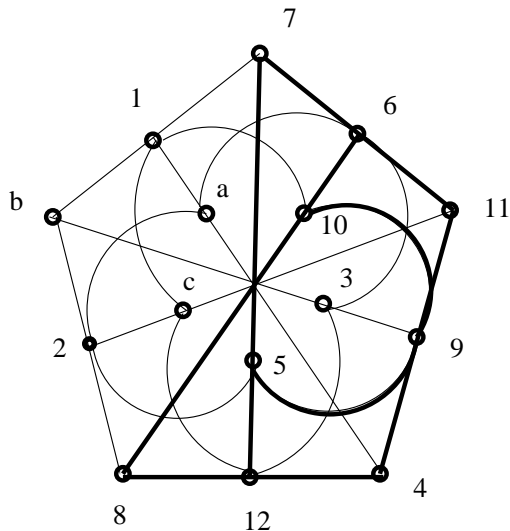}}
\caption{The generalized quadrangle $W(2)$ as the underlying geometry of two-qubit systems. The Pauli operators correspond to the points and maximally commuting subsets of them to the lines of the quadrangle.  Three operators on each line have a common base of eigenvectors; six out of fifteen such bases are entangled (the corresponding lines being indicated by boldfacing).}
\end{figure}

In this representation of the commutation relations it is also apparent that there is a six-fold symmetry of the two-qubit system [and of the generalized quadrangle $W(2)$]. Let us now see the quadrangle as a graph. Any point/vertex is connected to six distinct vertices and the whole graph can be defined as the complement $\hat{L}(K_6)$ of the line graph over the complete graph $K_6$ with six vertices. It is also known that the automorphism (symmetry) group of the quadrangle $W(2)$ is precisely the symmetric group with six elements $S(6)$. Finally there are six maximal sets of five disjoint lines (corresponding to six distinct complete sets of mutually unbiased bases) and any two distinct sets share a single line, a feature which can be still be represented as a complete graph $K_6$. It will be shown in Sec \ref{Cliff} that the six-fold symmetry also arises within the two-qubit Clifford group.

The geometry underlying multiple qubits is a symplectic polar space of order two \cite{Pauligraphs}. Totally isotropic subspaces, there are $(2+1)(2^2+1)...(2^n+1)$ of them within the polar space, correspond to mutually unbiased bases. See also Refs \cite{Ca97},\cite{Howe}. 

\section{The Clifford group}
\label{Cliff}

Let us assume a quantum computer in a state $\left|\psi\right\rangle$, and apply to it an error $g$ belonging to the Pauli group $\mathcal{P}$ so that the new state of the computer is $g\left|\psi\right\rangle $. One allows unitary evolutions $U$ so that the new state evolves as $ Ug\left|\psi\right\rangle=UgU^{\dag} U\left|\psi\right\rangle$. For stabilizing the error within the Pauli group $\mathcal{P}$, one requires that $UgU^{\dag} \in \mathcal{P}$. The set of operators which leaves $\mathcal{P}$ invariant under conjugation is the normalizer $\mathcal{C}_n$ in the unitary group $U(n)$, also known as the Clifford group $U(2^n)$ \cite{}. 

One learned from Gottesman-Knill theorem that the Hadamard  gate $H$ and the phase gate $P$ are in the one-qubit Clifford group $\mathcal{C}_1$, and that the controlled-$Z$ gate is in the two-qubit Clifford group $\mathcal{C}_2$. Any gate in $\mathcal{C}_n$ may be generated from the application of gates from $\mathcal{C}_1$ and $\mathcal{C}_2$ \cite{Go98,Clark07}. Clifford group $\mathcal{C}_n$ on $n$-qubits has order $|\mathcal{C}_n|=2^{n^2+2n+3}\prod_{j=1}^n 4^j-1$ \cite{Clark07}.

Below we will concentrate on the properties of the Clifford group related to one and two qubits, using the group theoretical package GAP4 \cite{GAP}. Generation of the gates will be ensured by the use of cyclotomic numbers, as described in Sec 18 of the GAP4 reference manual. For example the elements $1$, $-1$, $i$ and $2^{1/2}$ will be modelled as the roots of unity $E(1)$, $E(2)$ and $E(4)$ and as $ER(2)$, respectively. Some basic knowledge about finite group theory and group of automorphisms can be found in Ref \cite{Planat08}. 

\subsection{The Clifford group on a single qubit}
\label{single}

The one-qubit Clifford group is defined as $\mathcal{C}_1=\left\langle H,P \right\rangle$. It has order $|\mathcal{C}_1|=192$ and center isomorphic to the cyclic group $\mathcal{Z}_8$. One defines the {\it inner} group $G=\mathcal{C}_1/\mbox{Center}(\mathcal{C}_1)$. Elements $n$ of $G$ which are invariant under a conjugation $gng^{-1}=gng^{\dag}$, $\forall g \in G$, are in a normal subgroup $N$ of $G$. One only finds two non trivial normal subgroups of $G$. The first one $N_1$ is isomorphic to the Klein group $\mathcal{Z}_2 \times \mathcal{Z}_2$. The second one $N_2$ is isomorphic to the alternating group $A(4)$, the group of orientation preserving symmetries of the tetrahedron. Other useful relations are the isomorphisms $N_2/N_1\cong \mathcal{Z}_3$, $G/N_1 \cong S(3)$ so that $G \cong S(4) $, the symmetric group with four elements.


%
%

\subsection{The Clifford group on two qubits}
\label{two}

The two-qubit Clifford group may be generated as $\mathcal{C}_2=\left\langle H \otimes H,H \otimes P, CZ \right\rangle$. It has order $|\mathcal{C}_2|=92160$ and center isomorphic to $\mathcal{Z}_8$. The {\it inner} group $G=\mathcal{C}_2/\mbox{Center}(\mathcal{C}_2)$ is of order $11520$. 
As for the single qubit case, one only discovers two non trivial normal subgroups of $G$. The first one is $N_1 \cong \mathcal{Z}_2^{\times 4}$ and leads to an expression of the inner group as the semi-direct product $G=\mathcal{Z}_2^{\times 4} \rtimes S(6)$ \cite{Ca97}. The second normal subgroup $N_2=U_6\cong \mathcal{Z}_2^{\times 4}\rtimes A(6)$ is of order $5760$; it is a perfect group. It can be seen as a parent of the six element alternating group $A(6)$, because its outer automorphism group $\mbox{Out}(U_6)$ is the same, equal to the Klein group $\mathcal{Z}_2 \times \mathcal{Z}_2$. Useful expressions are $N_2/N_1 \cong A(6)$ and $G/N_1 \cong S(6)$. 

The group $U_6$ is an important maximal subgroup of several sporadic groups. The group of smallest size where it appears is the Mathieu group $M(22)$. 
Mathieu groups are sporadic simple groups, so that $U_6$ is not normal in $M(22)$. It appears in relation to a subgeometry of $M(22)$ known as an {\it hexad}. 
Let us recall the definition of Steiner systems. A Steiner system $S(a,b,c)$ with parameters $a$, $b$, $c$, is a $c$-element set together with a set of $b$-element subsets of $S$ (called {\it blocks}) with the property that each $a$-element subset of $S$ is contained in exactly one block. A finite projective plane of order q, with the lines as blocks, is an $S(2, q+1, q^2+q+1)$, since it has $q^2+q+1$ points, each line passes through $q+1$ points, and each pair of distinct points lies on exactly one line. Any large Mathieu group can be defined as the automorphism (symmetry) group of a Steiner system \cite{Wilson}. The group $M(22)$ stabilizes the Steiner system $S(3,6,22)$ comprising $22$ points and $6$ blocks, each set of $3$ points being contained exactly in one block\footnote{There exists up to equivalence a unique S(5,8,24) Steiner system called a Witt geometry. The group $M(24)$ is the automorphism group of this Steiner system, that is, the set of permutations which map every block to some other block. The subgroups $M(23)$ and $M(22)$ are defined to be the stabilizers of a single point and two points respectively.}. Any block in $S(3,6,22)$ is a Mathieu hexad, i.e. it is stabilized by the {\it general} alternating group $U_6$.   

\section{Discussion}

It is well known that among symmetric groups, only $S(6)$ has an outer automorphism, the group $\mathcal{Z}_2$. This also yields another outer automorphism attached to $A(6)$, the Klein group $\mathcal{Z}_2 \times \mathcal{Z}_2$. These exceptional automorphism groups among the symmetric and alternating groups may now be put in a geometrical perspective by their relationship to the hexads of the Mathieu group $M(22)$, and in a physical perspective by their relation ship to the two-qubit Pauli and Clifford groups.  

We believe that this research opens new vistas on several subjects and leaves open several questions. At the mathematical level, we would like to clarify the structure of higher order Clifford groups. It is not an easy computational task because already for three qubits the  {\it inner} group $G$ has  order $92 897 280$, which seems not to be the size of any known maximal subgroup of sporadic groups. At the quantum computational level, a challenge would be to relate the discovered $6$-fold symmetry of $\mathcal{C}_2$ to some clever gate design. Finally, the symmetries of $n$-qubit systems presumably relate to other branches of group theory such as Lie group theory, and algebra, such as Clifford algebra, insufficiently explored so far. See \cite{Planat08,PlanatKibler08,Planat09} for recent developments. 

As a final note, let us mention that the Klein group, which appears as the outer automorphism group of the alternating group $A(6)$ (and of its parent, the normal subgroup $U_6$ of the {\it inner} two-qubit Clifford group) can also be viewed as the additive group of the ring $\mathcal{R}=GF(2)\times GF(2)$. The projective line over $\mathcal{R}$ is a three by three grid (see the boldfaced lines in Fig. 1 which models the entanglement part in the quadrangle $W(2)$ and Ref \cite{Pauligraphs}). At least for the two-qubit system, there is a nice correspondence between the exceptional outer group of $A(6)$ and the so-called Mermin square used to prove the Kochen-Specker theorem in dimension four \cite{Mermin}. See also Ref \cite{Castro} for a different view related to Clifford spaces.

The Clifford group formalism may in principle be applied to any dimension \cite{Clark, Hostens}. It is another open problem to relate the corresponding Clifford groups to the symplectic polar spaces\cite{Pauligraphs} and ring projective lines attached to composite systems \cite{PlanatJPhysA}.

\section*{Acknowledgements}
The author acknowledges the support of the PEPS program (Projets Exploratoires Pluridisciplinaires) from the STIC department at CNRS, France (Sciences et Technologies de l'Information et de la Communication). The work was presented at the workshop {\it Prolegomena for quantum computing} held at FEMTO-ST, Besan\c{c}on, Nov 21-22, 2007. The author also acknowledges discussion with Metod Saniga at an early stage of the research.


\section*{Bibliography}

\vspace*{.0cm} \noindent
\vspace*{-.1cm}
\begin{thebibliography}{10}

\bibitem{Go98} The Heisenberg representation of quantum computers. D Gottesman. Preprint quant-ph/9807006.

\bibitem{Ca97} Quantum error correction and orthogonal geometry. A R Calderbank, E M Rains, P W Shor and N J A. Sloane. Phys Rev Lett. 78,405--408  (1997).    

\bibitem{Knill} Non-binary unitary error bases and quantum codes. E Knill. Preprint quant-ph/9608048. 

\bibitem{Kla1} Beyond stabilizer codes I: nice error bases. A Klappenecker and M. R\"{o}tteler. IEEE Trans Inform Theory 48, 2392--95 (2002).

\bibitem{Kla2} Beyond stabilizer codes II: Clifford codes. A Klappenecker and M. R\"{o}tteler. IEEE Trans Inform Theory 48, 2396--99 (2002).

\bibitem{Nielsen} Quantum computation and quantum information. M A Nielsen and I L Chuang. Cambridge Univ. Press (2000).

\bibitem{Raussen01} A one-way quantum computer. R Raussendorf and H Briegel. Phys Rev Lett 86, 5188-91 (2001).

\bibitem{Jorrand} Unifying quantum computation with projective measurements only and one way quantum computation. P. Jorrand and S. Perdrix. Preprint quant-ph/0404125.

\bibitem{Danos} The measurement calculus. V Danos, E Kashefi and P Panangaden. Preprint 0407.1263 [quant-ph].

\bibitem{GAP}
The GAP Group, GAP --- Groups, Algorithms, and Programming, Version 4.4; 2004. (http://www.gap-system.org).

\bibitem{Clark07} Generalized Clifford groups and simulation of associated quantum circuits. S Clark, R Jozsa and N Linden. Quant Inf Comp 8, 106--26 (2008). 

\bibitem{Appleby} Symmetric informationnaly complete-positive operator valued measures and the extended Clifford group. D M Appleby. J Math Phys 46, 521071--29 (2005).

\bibitem{Markham} Local encoding of classical information onto quantum states. Y Tanaka, D Markham and M Murao. Preprint quant-ph/0702190.

\bibitem{Pauligraphs}
On the Pauli graphs of $N$-qudits. M Planat and M Saniga. Quant Inf Comp 8, 127--46 (2008).

\bibitem{Mermin}
Hidden variables and two theorems of John Bell. N D Mermin. Rev Mod Phys 65, 803-15 (1993). 

\bibitem{PlanatMUBs}
A survey of finite algebraic geometrical structures underlying mutually unbiased measurements. M Planat, H C Rosu and S Perrine. Found of Phys 36, 1662--80 (2006).

\bibitem{Howe}
Nice error bases, mutually unbiased bases, induced representations, the Heisenberg group and finite geometries. R Howe. Indag Mathem, N S 16, 553--83 (2005).  

\bibitem{Planat08} 
On group theory for quantum gates and quantum coherence. Planat M and Jorrand P {\it J. Phys. A: Math. Theor.} {\bf 41} 182001 (2008).


\bibitem{PlanatKibler08}
 Unitary reflection groups for quantum fault tolerance. Planat M and Kibler M. Preprint 0807.3650 [quant-ph].

\bibitem{Wilson}
The finite simple groups. R A Wilson. available at http://www.maths.qmul.ac.uk/$\tilde{~}$raw/fsgs.html

\bibitem{Planat09} 
Clifford groups of quantum gates, BN-pairs and smooth cubic surfaces. Planat M and Sol\'e P. Preprint 0811.2109 [quant-ph]. 

\bibitem{Clark}
Valence bond solid formalism for $d$-level one-way quantum computation. S Clark. Preprint quant-ph/0512155.

\bibitem{Hostens}
Stabilizer states and Clifford operations for systems of arbitrary dimensions, and modular arithmetic. E Hostens, J Dehaene and B De Moor. Phys Rev A 71, 42315-24 (2005). 

\bibitem{Castro}
There is no Eistein-Podolski-Rosen paradox in Clifford spaces. C Castro. Adv Studies Theor Phys 1, 603-10 (2007).

\bibitem{PlanatJPhysA}
Qudits of composite dimension, mutually unbiased bases and projective ring geometry. M Planat and A C Baboin. J. Phys A Math and Theor 40, F1-F8 (2007).



\end{thebibliography}
\end{document}